\documentclass{article}
\usepackage{amssymb, amsmath}
\numberwithin{equation}{section}
\newtheorem{theorem}{Theorem}[section]
\newtheorem{proposition}{Proposition}[section]
\newtheorem{lemma}{Lemma}[section]
\def\supp{{\rm supp}}
\begin{document}
\title{On the non-relativistic limit of the spherically symmetric
Einstein-Vlasov-Maxwell system}
\author{P. Noundjeu \\
Department of Mathematics, Faculty of Science, University of
Yaounde 1,\\ PO Box 812, Yaounde, Cameroun \\ {e-mail:
noundjeu@uycdc.uninet.cm or pnoundjeu@yahoo.fr}}
\date{}
\maketitle
\begin{abstract}
The Einstein-Vlasov-Maxwell(EVM) system can be viewed as a
relativistic generalization of the Vlasov-Poisson(VP) system. As
it is proved below, one of nice property obeys by the first system
is that the strong energy condition holds and this allows to
conclude that the above system is physically viable. We show in
this paper that in the context of spherical symmetry, solutions of
the perturbed (EVM) system by $\gamma := 1/c^{2}$, $c$ being the
speed of light, exist and converge uniformly in the
$L^{\infty}$-norm, as $c$ goes to infinity on compact time
intervals to solutions of the non-relativistic (VP) system.
\end{abstract}
\section{Introduction}
The classical (VP) system models the time evolution of
collisionless particles in the Newtonian dynamic setting,
particles could be for instance atoms and molecules in neutral gas
or electrons and ions in a plasma. In stellar dynamics, particles
are either stars in galaxy or galaxies in a cluster of galaxies
\cite{andreasson}. The global behaviour of solutions of the above
system now is well understood (see \cite{lions},
\cite{pfaffelmoser}, \cite{rein1}, \cite{schaeffer1},
\cite{glassey}). In \cite{rein2}, the authors use an estimate to
prove that the Newtonian limit of the spherically symmetric
Vlasov-Einstein system is the classical (VP) system. This result
is extended by Rendall in \cite{rendall1} to the general
asymptotically flat Einstein-Vlasov system. The above
investigation concerns the case where the charge of particles is
small to be neglected.

We are inspired by what is done in \cite{rein2} and we want to
extend this result considering in this paper the (EVM) system that
models the time evolution of self-gravitating collisionless
charged particles in the general relativity setting. Firstly we
discuss and obtain that the above system satisfies the strong
energy condition, and then the system is physically viable.
Secondly, we perturb the (EVM) system by a parameter $\gamma =
1/c^{2}$ and together with the assumption of spherical symmetry,
we show using an estimate that for $\gamma$ small, the Cauchy
problem associated with the obtained system admits a unique
regular solution, with time existence interval independent of
$\gamma$. Once this result is obtained we deduce that if $c$ goes
to infinity then the above solution converges uniformly to a
solution of the classical (VP) system. To do so, we consider the
new constraint equations and discuss the existence of solutions
satisfying the constraints as we did in \cite{noundjeu1}. This
gives rise to a new mathematical feature since the present
equations contain two parameters $\gamma$ and $q$ ($q$ being the
charge of particle), rather than one as it is the case where
$\gamma = 1$. We obtain that for given initial datum to the
distribution function,  solutions of the constraints exist and
depend smoothly(i.e $C^{\infty}$) on parameter $\zeta := (\gamma,
q)$, for $\zeta$ small, and this allows us to construct the set of
initial data so that the initial datum for the metric function
$\lambda_{\gamma}$ is bounded in the $L^{\infty}$-norm. So, with
these initial data, we undertake the Cauchy problem for the
modified (EVM) system and establish as we did in \cite{noundjeu2},
the local existence and uniqueness theorem and continuation
criterion. The interest of this paper lies on the fact that the
estimates we use are complicated to establish. In fact, contrary
to the uncharged case, due to the presence of metric function
$\lambda_{\gamma}$ in the both side of equations, we have to
estimate both the supremum of momenta on the support of the
distribution function $f_{\gamma}$ and the supremum of
$e^{2\lambda_{\gamma}}$. We are not aware that the above is
already done.

It is appropriate at this point to put our investigation in the
context of general relativity. The classical limit of the
relativistic Vlasov-Maxwell system is studied by Jack Schaeffer
\cite{schaeffer2}. Using and estimate, the author proves that this
limit is the classical (VP) system. Also, the same result is
established by Asano \cite{asano} for the classical Vlasov-Maxwell
system. The recent result in this field is that of S. Calogero and
H.Lee \cite{calogero}. They consider the relativistic
Nordstr\"om-Vlasov system and establish that its non-relativistic
limit is the classical (VP) system.

The paper is organized as follows. In Sect.2, we recall the
classical (VP) system. In Sect.3, we introduce the general
formulation of the (EVM) system and we discuss the energy
conditions. In Sect.4, we introduce the perturbed spherically
symmetric (EVM) system and state the main results of this paper.
\section{The classical(non-relativistic) (VP) system}
In the Newtonian dynamic, the time evolution of a set of
collisionless particles is governed by the following equations
known as the (VP) system:
\begin{equation} \label{eq:2.1}
\partial_{t} f + v \cdot \nabla_{\tilde{x}}f - \nabla_{\tilde{x}}U
\cdot \nabla_{\tilde{p}}f = 0
\end{equation}
\begin{equation} \label{eq:2.2}
 \nabla_{\tilde{x}}U = 4\pi \eta M, \quad \eta = \pm 1, \quad M :=
 \int_{\mathbb{R}^{3}}f(t, \tilde{x}, \tilde{p})d\tilde{p}
\cdot \nabla_{\tilde{p}}f = 0
\end{equation}
Here (\ref{eq:2.1}) is the Vlasov equation for the unknown $f$,
$f$ being the distribution function that measures the probability
density to find a particle (star) at time $t$ with position
$\tilde{x}$ and with momentum $\tilde{p}$, where $t > 0$,
$(\tilde{x}, \tilde{p}) \in \mathbb{R}^{2}$. Note that $f$ is
defined on the mass shell. (\ref{eq:2.2}) is the Poisson equation
for the unknown $U = U(t, \tilde{x})$ that measures the Newtonian
potential generated by stars. If $\eta = -1$ then (\ref{eq:2.1})
and (\ref{eq:2.2}) model the plasma physics case. In what follows,
we consider the case where $\eta = 1$.
\section{The (EVM) system}
As we said before, we take fast moving collisionless particles
with unit mass and charge $q$. The gravitational constant and the
speed of light are taken to be equal to unity. The basic spacetime
is $(\mathbb{R}^{4}, g)$, with $g$ a Lorentzian metric with
signature $(-, +, +, +)$. In the sequel, we assume that Greek
indices run from $0$ to $3$ and Latin indices from $1$ to $3$. We
adopt the Einstein summation convention. The (EVM) system reads in
local coordinates at a given point $(x^{\alpha}) = (t,
\tilde{x})$:
\begin{equation} \label{eq:3.1}
R_{\alpha \beta} - \frac{1}{2}g_{\alpha \beta}R = 8\pi (T_{\alpha
\beta}(f) + \tau_{\alpha \beta}(F))
\end{equation}
\begin{equation} \label{eq:3.2}
\frac{\partial f}{\partial t} + \frac{p^{i}}{p^{0}}\frac{\partial
f}{\partial x^{i}} - (\Gamma_{\beta \tau}^{i}p^{\beta}p^{\tau} +
qp^{\beta}F_{\beta}^{\, \, \, i})\frac{\partial f}{\partial p^{i}}
= 0
\end{equation}
\begin{equation} \label{eq:3.3}
\nabla_{\alpha}F^{\alpha \beta} = J^{\beta}(f); \quad
\nabla_{\alpha}F_{\beta \gamma} + \nabla_{\beta}F_{\gamma \alpha}
+ \nabla_{\gamma}F_{\alpha \beta} = 0,
\end{equation}
with:
\begin{align*}
T_{\alpha \beta}(f) &= - \int_{\mathbb{R}^{3}} p_{\alpha}
p_{\beta} \omega_{p}; \quad \tau_{\alpha \beta} = -
\frac{g_{\alpha \beta}}{4}F_{\gamma \nu}F^{\gamma \nu} + F_{\beta
\gamma}F_{\alpha}^{\, \, \, \gamma}\\
J^{\beta}(f)(x) &= q \int_{\mathbb{R}^{3}}p^{\beta}f(x,
p)\omega_{p}; \, \omega_{p} = \mid g
\mid^{1/2}\frac{dp^{1}dp^{2}dp^{3}}{p_{0}}, \, p_{0} = g_{0 0}
p^{0},
\end{align*}
where $\Gamma_{\lambda \mu}^{\alpha}$ denote the Christofell
symbols. In the above, $f$ stands for the distribution function of
the charged particles defined on the mass shell:
\begin{equation*}
g_{\alpha \beta}p^{\alpha}p^{\beta} = - 1,
\end{equation*}
$F$ stands for the electromagnetic field created by the charged
particles. Here (\ref{eq:3.1}) is the Einstein equations for the
metric tensor $g = (g_{\alpha \beta})$ with sources generated by
both $f$ and $F$, that appear in the stress-energy tensor
$T_{\alpha \beta} + \tau_{\alpha \beta}$. Equation (\ref{eq:3.2})
is the Vlasov equation for $f$ and (\ref{eq:3.3}) are the Maxwell
equations.
\subsection{The energy conditions}
Physically, the quantity $(T_{\alpha \beta} + \tau_{\alpha
\beta})V^{\alpha}V^{\beta}$ represents the energy density of
charged particles obtained by an observer whose $4$-velocity is
$(V^{\alpha})$. So, for any physically viable theory, this
quantity is nonnegative for every timelike vector $(V^{\alpha})$,
and the above assumption is known as the weak energy condition. We
are going to show that in fact, the dominant energy condition
holds, i.e: $(T_{\alpha \beta} + \tau_{\alpha
\beta})V^{\alpha}W^{\beta} \geq 0$, for all future-pointing
timelike vectors $(V)^{\alpha}$ and $(W^{\alpha})$. This implies
the weak energy condition. Note also that the above definition of
dominant energy condition is equivalent to that given by Hawking
and Ellis in [\cite{hawking}, p.91]. Once this is clarified, we
show that since the Maxwell tensor is traceless, and the strong
energy condition that is $R_{\alpha \beta}V^{\alpha}V^{\beta} \geq
0$, for every timelike vector $(V^{\alpha})$, holds for the
Einstein-Vlasov system, the same is true in our context.
\begin{lemma} \label{L:3.1}
The following assertions are equivalent:
\begin{itemize}
\item[1)] For any two future-pointing timelike vectors
$(V^{\alpha})$, $(W^{\alpha})$, one has: \\ $(T_{\alpha \beta} +
\tau_{\alpha \beta})V^{\alpha}W^{\beta} \geq 0$
\item[2)] For every timelike vector $(V^{\alpha})$, one has: $(T_{\alpha \beta} +
\tau_{\alpha \beta})V^{\alpha}V^{\beta} \geq 0$, and $(T_{\alpha
\beta} + \tau_{\alpha \beta})V^{\beta}$ is a non-spacelike vector.
\end{itemize}
\end{lemma}
\textbf{Proof:} We first prove that 2) implies 1). If
$(V^{\alpha})$ and $(W^{\alpha})$ are two future-pointing timelike
vectors then 2) implies that $(T_{\alpha \beta} + \tau_{\alpha
\beta})V^{\beta}$ is non-spacelike. This means by definition that
it is either timelike or null. Using once again 2), its
contraction with $(V^{\alpha})$ is positive. Thus, it is in fact
past-pointing timelike or null, its opposite is future-pointing
timelike and then its contraction with $(W^{\alpha})$ is positive.
This comes from the fact that a contraction of two future-pointing
timelike vectors is negative or null.  So 1) is proved.

Now conversely, suppose that 1) holds and let us prove 2). Let
$(V^{\alpha})$ be a timelike vector. If $V^{0} > 0$, then
$(V^{\alpha})$ is future-pointing timelike and the first condition
in 2) holds by taking $V = W$. If $V^{0} < 0$, then $(-V^{0},
-V^{i})$ is future-pointing timelike and we can conclude as we
made before for the first condition in 2). For the second part,
suppose that $(V^{\alpha})$ is future-pointing timelike and define
$P_{\alpha} := (T_{\alpha \beta} + \tau_{\alpha \beta})V^{\beta}$.
Condition 1) implies that $(P_{\alpha})$ satisfies $P_{\alpha}
W^{\alpha} \geq 0$ for every future-pointing timelike vector
$(W^{\alpha})$. We aim to show that $(P_{\alpha})$ is
non-spacelike. To do this, let us assume that $(P_{\alpha})$ is
spacelike. and get a contradiction. Set $L := P_{\alpha}
P^{\alpha}$. By assumption $L > 0$. Let $(T^{\alpha})$ be a
future-pointing timelike vector orthogonal to $(P_{\alpha})$ with
$T_{\alpha} T^{\alpha} = -L$ and $T^{0} > P^{0}$(for the
construction of vector $(T^{\alpha})$, one can take for instance
in normal coordinates: $T^{0} = \sqrt{\overset{3}{\underset{i =
1}{\sum}}(P^{i})^{2}}$, $T^{i} = P^{0}P^{i}/T^{0}$). Set
$W^{\alpha} = 2T^{\alpha} - P^{\alpha}$. Then
$W^{\alpha}W_{\alpha} = -3L < 0$, $W^{0} = 2T^{0} - P^{0} > 0$ and
$(W^{\alpha})$ is a future-pointing timelike vector, and
$W^{\alpha} P_{\alpha} = -L < 0$. This is the desired
contradiction. Now if $(V^{\alpha})$ is past-pointing timelike,
then $(-V^{\alpha})$ is future-pointing and follow the first step
of the proof in which $P_{\alpha}$ is replaced by $-P_{\alpha} =
(T_{\alpha \beta} + \tau_{\alpha \beta})(-V^{\beta})$.
Analogously, we are led to $(-P_{\alpha})$ is non-spacelike and so
is $(P_{\alpha})$. In conclusion, the second part of condition 2)
holds, for every timelike vector $(V^{\alpha})$ and the proof is
complete.
\begin{proposition} \label{P:3.1}
\begin{itemize}
\item[1)] For every two future-pointing vectors $(V^{\alpha})$,
$(W^{\alpha})$, one has:
\begin{equation} \label{eq:3.4}
T_{\alpha \beta}V^{\alpha}W^{\beta} + \tau_{\alpha
\beta}V^{\alpha}W^{\beta} \geq 0
\end{equation}
\item[2)] For every timelike vector $(V^{\alpha})$, one has:
\begin{equation*}
R_{\alpha \beta}V^{\alpha}V^{\beta} \geq 0.
\end{equation*}
\end{itemize}
\end{proposition}
\textbf{Proof:} Consider part 1) of the above proposition. Since
Penrose and Rindler state the dominant energy condition for the
Maxwell equations in writing the Maxwell tensor $\tau_{\alpha
\beta}$ as a quadratic form of spinor fields in \cite{penrose},
the second term in the left hand side of (\ref{eq:3.4}) is
nonnegative and we just need to establish the same result for the
first term in the left hand side of (\ref{eq:3.4}). Let
$(V^{\alpha})$, $(W^{\alpha})$ be two future-pointing timelike
vectors. Taking the first term in the left hand side of
(\ref{eq:3.4}), we obtain, since $f \geq 0$, $(p^{\alpha})$ is
future-pointing timelike vector and $-p_{0} > 0$, one has:
\begin{equation*}
T_{\alpha \beta}V^{\alpha}W^{\beta} =
\int_{\mathbb{R}^{3}}(p_{\alpha}V^{\alpha})(p_{\beta}W^{\beta})f
\mid g \mid^{\frac{1}{2}}\frac{dp^{1}dp^{2}dp^{3}}{-p_{0}} \geq 0
\end{equation*}
So, (\ref{eq:3.4}) holds as announced. Now concerning part 2) of
the above Proposition, the contraction of (\ref{eq:3.1}) gives,
since $g^{\alpha \beta}\tau_{\alpha \beta} = 0$:
\begin{equation*}
R = -8\pi T
\end{equation*}
where $T := g^{\alpha \beta}T_{\alpha \beta}$. Insertion of the
above in (\ref{eq:3.1}) yields:
\begin{equation*}
R_{\alpha \beta} = - 4\pi Tg_{\alpha \beta} + 8\pi (T_{\alpha
\beta} + \tau_{\alpha \beta}).
\end{equation*}
Next, let $(V^{\alpha})$ be a timelike vector. Then
\begin{equation} \label{eq:3.5}
R_{\alpha \beta}V^{\alpha}V^{\beta} = 8\pi \tau_{\alpha
\beta}V^{\alpha}V^{\beta} + 8\pi \left( T_{\alpha \beta} -
\frac{1}{2}Tg_{\alpha \beta} \right)V^{\alpha}V^{\beta}
\end{equation}
Since the Maxwell tensor satisfies the dominant energy condition
and then the weak energy condition, we can deduce that the first
term in the right hand side of (\ref{eq:3.5}) is nonnegative.
Also, the strong energy condition holds for the Einstein-Vlasov
system (for more details one can refer to [\cite{rendall2},
p.37-38]). The latter shows that the second term in the right hand
side of (\ref{eq:3.5}) is nonnegative and the strong energy
condition holds in our context.
\section{The perturbed spherically symmetric (EVM) system and the main results}
We consider the Lorentzian spacetime $(\mathbb{R}^{4},
g_{\gamma})$ where $g_{\gamma}$ is obtained by scaling $g$ with
$\gamma = 1/c^{2}$ in the spatial part. With the assumption of
spherical symmetry, we take $g_{\gamma}$ of the following form:
\begin{equation*}
ds^{2} = - e^{2\mu}dt^{2} + \gamma e^{2\lambda}dr^{2} +
r^{2}\gamma (d\theta^{2} + \sin^{2}\theta d\varphi^{2})
\end{equation*}
where $\mu = \mu(t, r)$; $\lambda = \lambda(t, r)$; $t \in
\mathbb{R}$; $r \in [0, + \infty[$; $\theta \in [0, \pi]$;
$\varphi \in [0, 2\pi]$.

Next, the assumption of spherical symmetry allows us to deduce the
following perturbed (EVM) system in the $(t, \tilde{x}, v)$
coordinates (see \cite{noundjeu3}); where
\begin{equation*}
v^{i} := p^{i} + (e^{\lambda} - 1)\frac{\tilde{x} \cdot
\tilde{p}}{r}\frac{x^{i}}{r}:
\end{equation*}
\begin{equation} \label{eq:4.1}
e^{-2\lambda}(2r\lambda' - 1) + 1 = 8\pi \gamma r^{2}\rho
\end{equation}
\begin{equation} \label{eq:4.2}
e^{-2\lambda}(2r\mu' - 1) + 1 = 8\pi \gamma^{2} r^{2}p
\end{equation}
\begin{equation} \label{eq:4.3}
\frac{\partial f}{\partial t} + e^{\mu - \lambda} \frac{v}{\sqrt{1
+ \gamma v^{2}}} \cdot \frac{\partial f}{\partial \tilde{x}} -
\left( e^{\mu - \lambda} \mu' \frac{1}{\gamma}\sqrt{1 + \gamma
v^{2}} + \dot{\lambda} \frac{\tilde{x}.v}{r} - q e^{\lambda + \mu}
e(t, r) \right)\frac{\tilde{x}}{r} \cdot \frac{\partial
f}{\partial v} = 0
\end{equation}
\begin{equation} \label{eq:4.4}
\frac{\partial}{\partial r}(r^{2}e^{\lambda}e) =
qr^{2}\gamma^{3/2}e^{\lambda}M
\end{equation}
where $\lambda' = \frac{d\lambda}{dr}$; \, $\dot{\lambda} =
\frac{d\lambda}{dt}$ and:
\begin{equation} \label{eq:4.5}
\rho(t, \tilde{x}) := \int_{\mathbb{R}^{3}} f(t,
\tilde{x},v)\sqrt{1 + \gamma v^{2}} dv + \frac{1}{2}
e^{2\lambda(t, \tilde{x})} e^{2}(t, \tilde{x})
\end{equation}
\begin{equation} \label{eq:4.6}
p(t,\tilde{x}) := \int_{\mathbb{R}^{3}} \left(
\frac{\tilde{x}.v}{r} \right)^{2} f(t, \tilde{x}, v)
\frac{dv}{\sqrt{1 + \gamma v^{2}}} - \frac{1}{2} e^{2\lambda(t,
\tilde{x})} e^{2}(t, \tilde{x})
\end{equation}
\begin{equation} \label{eq:4.7}
M(t, \tilde{x}) := \int_{\mathbb{R}^{3}} f(t, \tilde{x}, v)dv.
\end{equation}
Next, it is well known that in the context of spherical symmetry,
the (VP) system reduces to:
\begin{equation} \label{eq:4.8}
\partial_{t}f + v \cdot \partial_{\tilde{x}}f - K(t, \tilde{x})
\cdot \partial_{v}f = 0
\end{equation}
\begin{equation} \label{eq:4.9}
K(t, \tilde{x}) = \frac{1}{r^{2}}\frac{\tilde{x}}{r}\int_{\mid y
\mid \leq r} M(t, y)dy.
\end{equation}
In the sequel we denote by $(EVM_{\gamma})$ the system
(\ref{eq:4.1}), (\ref{eq:4.2}), (\ref{eq:4.3}) and (\ref{eq:4.4}).
It is prescribed on $(EVM_{\gamma})$ the initial data $\lambda(0)
= \overset{\circ}{\lambda}$ and $f(0) = \overset{\circ}{f}$, where
$\overset{\circ}{f}$ is nonnegative, spherically symmetric
function i.e invariant when one applies any rotation of
$\mathbb{R}^{3}$ on both the variables $\tilde{x}$ and $v$,
$\overset{\circ}{\lambda}$ being a given function. Concerning the
system (\ref{eq:4.8}) and (\ref{eq:4.9}), only the initial
condition $f(0) = \overset{\circ}{f}$ is needed. Also, we are
interesting with the asymptotic flatness of spacetime and a
regular center at $r = 0$. So, we have the following boundary
conditions:
\begin{equation} \label{eq:4.10}
\lim_{r \to + \infty}\lambda(t, r) = \lim_{r \to + \infty}\mu(t,
r) = \lambda(0, r) = 0, \quad t \geq 0.
\end{equation}
Now, the assumption that the electric field $E$ defined by $E(t,
\tilde{x}) = e(t, r)\frac{\tilde{x}}{r}$, is spherically symmetric
and the fact that at the spatial infinity there is no charged
particle lead to the following  boundary condition (for more
details, see \cite{noundjeu3}):
\begin{equation} \label{eq:4.11}
\lim_{r \to + \infty}e(t, r) = e(0, r) = 0, \quad t \geq 0.
\end{equation}
Before stating the main result of this section, we recall the
following result whose proof is made by straightforward
calculation (see \cite{noundjeu3}). The concept of regularity of
solutions we use is the same as in \cite{noundjeu3}.
\begin{lemma} \label{L:4.1}
Take $\gamma = 1$. Let $(f, \lambda, \mu, e)$ be a regular
solution of $EVM_{1}$ satisfying (\ref{eq:4.10}) and
(\ref{eq:4.11}) on some interval $I$. Then for every $a \in
\mathbb{R} \setminus \{ 0 \}$,
\begin{equation*}
\begin{cases}
f_{a}(t, \tilde{x}, v) := a^{2}f(at, a\tilde{x}, v); \quad
\lambda_{a}(t, r) := \lambda(at, ar)\\
\mu_{a}(t, r) := \mu(at, ar); \quad e_{a}(t, r) := ae(at, ar)
\end{cases}
\end{equation*}
defines another regular solution of this system on the interval
$a^{-1}I$.
\end{lemma}
Another thing to discuss is the existence of initial data
satisfying the constraints. The constraint equations are obtained
by setting $t = 0$ in (\ref{eq:4.1}), (\ref{eq:4.2}) and
(\ref{eq:4.4}). As we said in \cite{noundjeu1}, these constraints
reduce to
\begin{equation} \label{eq:4.12}
e^{-2\overset{\circ}{\lambda}}(2r\overset{\circ}{\lambda}' - 1) +
1 = 8\pi \gamma r^{2}\overset{\circ}{\rho}
\end{equation}
\begin{equation} \label{eq:4.13}
\frac{d}{dr}(r^{2}e^{\overset{\circ}{\lambda}}\overset{\circ}{e})
= q r^{2} \gamma^{3/2} e^{\overset{\circ}{\lambda}}
\overset{\circ}{M}
\end{equation}
where
\begin{equation*}
\overset{\circ}{e}(r) := e(0, r); \quad \overset{\circ}{\rho}(r)
:= \rho(0, r); \quad \overset{\circ}{M}(r) := M(0, r)
\end{equation*}
If $\gamma \in ]0, 1]$ and
\begin{equation} \label{eq:4.14}
8\pi \int_{0}^{r} s^{2} ds \int_{\mathbb{R}^{3}}
\overset{\circ}{f}(s,v) \sqrt{1 + v^{2}} dv < r, \quad r > 0,
\end{equation}
then
\begin{equation*}
8\pi \gamma \int_{0}^{r} s^{2} ds \int_{\mathbb{R}^{3}}
\overset{\circ}{f}(s,v) \sqrt{1 + \gamma v^{2}} dv < r, \quad r >
0.
\end{equation*}
So, under the assumption (\ref{eq:4.14}), we can apply the results
obtained in \cite{noundjeu1} to (\ref{eq:4.12}) and
(\ref{eq:4.13}), and deduce the following
\begin{proposition} \label{P:4.1}
Let $ \overset{\circ}{f} \in C^{\infty}(\mathbb{R}^{6})$ be
nonnegative, compactly supported, spherically symmetric and
satisfies (\ref{eq:4.14}). Take $\gamma \in ]0, 1]$. Then, for
$\zeta = (\gamma, q)$ small enough, the equations (\ref{eq:4.12})
and (\ref{eq:4.13}) have a unique global solution
$(\overset{\circ}{\lambda}_{\gamma}, \overset{\circ}{e}_{\gamma})
\in (C^{\infty}([0, + \infty[))^{2}$ such that
$\overset{\circ}{\lambda}_{\gamma}(0) =
\overset{\circ}{e}_{\gamma}(0) = 0$. Moreover, this solution
depends smoothly (i.e $C^{\infty}$) on the parameter $\zeta$.
\end{proposition}
Next, with the help of Proposition 4.1, for $\gamma \in ]0, 1]$
and $\Lambda > 0$, one can define the set of initial data as:
\begin{align*}
&D := \{ (\overset{\circ}{f}, \overset{\circ}{\lambda}_{\gamma},
\overset{\circ}{e}_{\gamma}) \in C^{\infty}(\mathbb{R}^{6}) \times
(C^{1}([0, + \infty[))^{2}, \quad \overset{\circ}{f} \geq 0,
\text{spherically symmetric},\\
& \qquad \text{and satisfies} \quad (\ref{eq:4.14}) \quad
\text{and} \quad
(\overset{\circ}{\lambda}_{\gamma},\overset{\circ}{e}_{\gamma})\\
& \qquad \text{is a regular solution of the constraints with} \,
\parallel \overset{\circ}{\lambda}_{\gamma} \parallel_{L^{\infty}}
\leq \Lambda \}.
\end{align*}
We now study the Cauchy problem given by $(EVM_{\gamma})$, the
initial data, (\ref{eq:4.10}) and (\ref{eq:4.11}).
\begin{theorem} \label{T:4.1}
There exists $T > 0$ and a continuous function $u : [0, T[
\rightarrow \mathbb{R}_{+}$ such that for $\gamma \in ]0, 1]$ and
$(\overset{\circ}{f}, \overset{\circ}{\lambda}_{\gamma},
\overset{\circ}{e}_{\gamma}) \in D$, the system $(EVM_{\gamma})$
has a unique regular solution $(f_{\gamma}, \lambda_{\gamma},
\mu_{\gamma}, e_{\gamma})$ on the interval $[0, T[$ with initial
data $(\overset{\circ}{f}, \overset{\circ}{\lambda}_{\gamma},
\overset{\circ}{\mu}_{\gamma}, \overset{\circ}{e}_{\gamma})$ and
\begin{equation*}
f_{\gamma}(t, \tilde{x}, v) = 0, \quad \mid v \mid > u(t), \quad
\tilde{x} \in \mathbb{R}^{3}, t \in [0, T[,
\end{equation*}
where $\overset{\circ}{\mu}_{\gamma} := \mu_{\gamma}(0, r)$.
\end{theorem}
\textbf{Proof:} Take $\gamma \in ]0, 1]$ and $(\overset{\circ}{f},
\overset{\circ}{\lambda}_{\gamma}, \overset{\circ}{e}_{\gamma})
\in D$. For $\gamma = 1$, it is shown in \cite{noundjeu2} that the
system $(EVM_{1})$ admits a unique regular local solution $(f,
\lambda, \mu, e)$ defined on a maximal existence interval $[0,
T_{1}[$, with initial data $(\overset{\circ}{f},
\overset{\circ}{\lambda}_{1}, \overset{\circ}{\mu}_{1},
\overset{\circ}{e}_{1})$, and using Lemma 4.1, one deduces that
$(f_{\gamma} := \gamma^{3/2} f(\gamma^{1/2}., \gamma^{1/2}., .),
\lambda_{\gamma} := \lambda(\gamma^{1/2}., \gamma^{1/2}.),
\mu_{\gamma} := \mu(\gamma^{1/2}., \gamma^{1/2}.), e_{\gamma} :=
\gamma^{1/2} e(\gamma^{1/2}., \gamma^{1/2}.))$ solves
$(EVM_{\gamma})$ on $[0, T_{\gamma}[$, where $T_{\gamma} :=
\gamma^{-1/2}T_{1}$. Set for every $t \in [0, T_{\gamma}[$
\begin{align*}
U_{\gamma}(t) := \sup \{ \mid v \mid \, | (\tilde{x}, v) \in \supp
f_{\gamma}(t) \}\\
Q_{\gamma}(t) := \sup \{ e^{2\lambda_{\gamma}(t, r)} \, | \, r
\geq 0\}.
\end{align*}
We now look for an estimate for $U_{\gamma}$ and $Q_{\gamma}$. Let
$s \mapsto (X_{\gamma}(s, t, \tilde{x}, v), V_{\gamma}(s, t,
\tilde{x}, v))$ be the solution of characteristic system
\begin{align*}
\dot{\tilde{x}} &= e^{\mu_{\gamma} -
\lambda_{\gamma}}\frac{v}{\sqrt{1 + \gamma v^{2}}}\\
\dot{v} &= - \left( \dot{\lambda}_{\gamma}\frac{\tilde{x} \cdot
v}{r} + e^{\mu_{\gamma} -
\lambda_{\gamma}}\frac{1}{\gamma}\mu_{\gamma}'\sqrt{1 + \gamma
v^{2}} - qe^{\mu_{\gamma} + \lambda_{\gamma}}e_{\gamma}
\right)\frac{\tilde{x}}{r}
\end{align*}
with $X_{\gamma}(t, t, \tilde{x}, v) = x$ and $V_{\gamma}(t, t,
\tilde{x}, v)$ = v. It is well known that the solution of
(\ref{eq:4.3}) with initial datum $\overset{\circ}{f}$ is given by
\begin{equation*}
f_{\gamma} (t, \tilde{x}, v) = \overset{\circ}{f}(X_{\gamma}(0, t,
\tilde{x}, v), V_{\gamma}(0, t, \tilde{x}, v))
\end{equation*}
and $\parallel f_{\gamma}(t) \parallel_{L^{\infty}} = \parallel
\overset{\circ}{f} \parallel_{L^{\infty}}$, $t \geq 0$. Along a
characteristic, we obtain since $\mu_{\gamma} - \lambda_{\gamma}
\leq 0$: $\mid  \dot{\tilde{x}} \mid \leq \mid v \mid / \sqrt{
\gamma v^{2}} \leq \frac{1}{\sqrt{\gamma}}$ and one deduces from
the above by integration on $[0, t]$ that $r := \mid \tilde{x}
\mid \leq r_{0} + \frac{t}{\sqrt{\gamma}}$, where
\begin{equation*}
r_{0} := \sup \{ \mid \tilde{x} \mid \, |\, (\tilde{x}, v) \in
\supp \overset{\circ}{f} \}
\end{equation*}
The integration of (\ref{eq:4.4}) yields:
\begin{equation} \label{eq:4.15}
e_{\gamma} (t, r) =
\frac{q}{r^{2}}\gamma^{3/2}e^{-\lambda_{\gamma}(t,
r)}\int_{0}^{r}s^{2}e^{\lambda_{\gamma}(t, s)}M_{\gamma}(t, s)ds
\end{equation}
where $M_{\gamma}$ is deduced from (\ref{eq:4.7}), replacing $f$
by $f_{\gamma}$. Now, since $-\lambda_{\gamma} \leq 0$,
distinguishing the cases $r < r_{0} + \frac{t}{\sqrt{\gamma}}$ and
$r \geq r_{0} + \frac{t}{\sqrt{\gamma}}$, one obtains the
following estimate for $e_{\gamma}$, since $\gamma \in ]0, 1]$:
\begin{equation} \label{eq:4.16}
\mid e_{\gamma} (t, r) \mid \leq C (1 + r_{0} +
t)Q_{\gamma}^{1/2}(t)U_{\gamma}^{3}(t), \quad r \geq 0
\end{equation}
where $C = C(q, \parallel \overset{\circ}{f}
\parallel_{L^{\infty}})$ is a constant. Using (\ref{eq:4.16}), we
deduce
\begin{equation} \label{eq:4.17}
(e^{2\lambda_{\gamma}}e_{\gamma}^{2})(t, r) \leq C(1 + r_{0} +
t)^{2}Q_{\gamma}^{3/2}(t)U_{\gamma}^{6}(t), \quad r \geq 0
\end{equation}
We use (\ref{eq:4.17}) and estimates obtained in the proof of
Theorem 1 in \cite{rein2} to deduce and estimates for
$\rho_{\gamma}$ and $p_{\gamma}$:
\begin{align*}
\parallel \rho_{\gamma}(t) \parallel_{L^{\infty}} &\leq
CU_{\gamma}^{3}(t)\sqrt{1 + U_{\gamma}^{2}(t)} + C(1 + r_{0} +
t)^{2}Q_{\gamma}^{3/2}(t)U_{\gamma}^{6}(t), \\
\parallel p_{\gamma}(t) \parallel_{L^{\infty}} &\leq
C\min \left\{ \frac{1}{\sqrt{\gamma}}U_{\gamma}^{4}(t),
U_{\gamma}^{5}(t) \right\} + C(1 + r_{0} +
t)^{2}Q_{\gamma}^{3/2}(t)U_{\gamma}^{6}(t).
\end{align*}
Here $\rho_{\gamma}$ and $p_{\gamma}$ are deduced from
(\ref{eq:4.5}) and (\ref{eq:4.6}) replacing $f$, $\lambda$ and $e$
by $f_{\gamma}$, $\lambda_{\gamma}$ and $e_{\gamma}$ respectively.
We now introduce the mass function given by:
\begin{equation*}
m_{\gamma}(t, r) := 4\pi \int_{0}^{r}s^{2}\rho_{\gamma}(t, s)ds =
\int_{\mid y \mid \leq r}\rho_{\gamma}(t, y)dy
\end{equation*}
and with the help of the estimate of $\rho_{\gamma}$, one obtains,
after distinguishing the cases $r \leq r_{0} + \frac{t}{\sqrt{t}}$
and $r > r_{0} + \frac{t}{\sqrt{\gamma}}$, and using the relation
\begin{equation*}
(e^{\lambda_{\gamma}}e_{\gamma})(t, r) = \left( \frac{r_{0} +
t/\sqrt{\gamma}}{r} \right)^{2} (e^{\lambda_{\gamma}} e_{\gamma})(
t, r_{0} + t/\sqrt{\gamma}), \quad r \in [r_{0} + t/\sqrt{\gamma},
+ \infty[
\end{equation*}
Now, one has:
\begin{align*}
\frac{m_{\gamma}(t, r)}{r} &\leq 4\pi \int_{0}^{r}
s\rho_{\gamma}(t, s)ds\\
&\leq C\left( r_{0} + \frac{t}{\sqrt{\gamma}}
\right)^{2}U_{\gamma}^{3}(t)\sqrt{1 + U_{\gamma}^{2}(t)} + C\left(
r_{0} + \frac{t}{\sqrt{\gamma}} \right)^{2}(1 + r_{0} +
t)^{2}Q_{\gamma}^{3/2}U_{\gamma}^{6}(t)\\
\frac{m_{\gamma}(t, r)}{r^{2}} &\leq C\left( r_{0} +
\frac{t}{\sqrt{\gamma}} \right)U_{\gamma}^{3}(t)\sqrt{1 +
U_{\gamma}^{2}(t)} + C\left( r_{0} + \frac{t}{\sqrt{\gamma}}
\right)(1 + r_{0} + t)^{2}Q_{\gamma}^{3/2}U_{\gamma}^{6}(t)
\end{align*}
The integration of (\ref{eq:4.1}) on $[0, r]$ yields
$e^{-2\lambda_{\gamma}}$ as a function of $m_{\gamma}$, and the
insertion of this relation in (\ref{eq:4.2}) gives
$\mu_{\gamma}'$. So multiplying the obtained equation by
$e^{\mu_{\gamma} - \lambda_{\gamma}}$ yields:
\begin{equation*}
(e^{\mu_{\gamma} - \lambda_{\gamma}}\mu_{\gamma}')(t, r) =
e^{\mu_{\gamma} + \lambda_{\gamma}}\left( \gamma
\frac{m_{\gamma}(t, r)}{r^{2}} + 4\pi \gamma^{2}rp_{\gamma}(t, r)
\right).
\end{equation*}
Since $\mu_{\gamma} + \lambda_{\gamma} \leq 1$, one obtains the
following estimate, after distinguishing the cases $r \leq r_{0} +
t/\sqrt{\gamma}$ and $r > r_{0} + t/\sqrt{\gamma}$ for the second
term in the right hand side of the expression above:
\begin{equation} \label{eq:4.18}
\begin{aligned}
\left| \frac{1}{\gamma}e^{\mu_{\gamma} - \lambda_{\gamma}}
\mu_{\gamma}' \right| &\leq C(1 + r_{0} +
t)U_{\gamma}^{3}(t)\sqrt{1 + U_{\gamma}^{2}(t)} + C(1 + r_{0} +
t)^{3}Q_{\gamma}^{3/2}(t)U_{\gamma}^{6}(t)\\
& \qquad + C(1 + r_{0} + t)U_{\gamma}^{5}(t) + (1 + r_{0} +
t)^{3}Q_{\gamma}^{2}(t)U_{\gamma}^{6}(t).
\end{aligned}
\end{equation}
Another part of the Einstein equations is (see \cite{noundjeu2}):
\begin{equation*}
\dot{\lambda}_{\gamma}(t, r) = -4\pi \gamma re^{\mu_{\gamma} +
\lambda_{\gamma}}k_{\gamma}(t, r),
\end{equation*}
where
\begin{equation*}
k_{\gamma}(t, r) = k_{\gamma}(t, \tilde{x}):=
\int_{\mathbb{R}^{3}}\frac{\tilde{x} . v}{r}f_{\gamma}(t,
\tilde{x}, v)dv,
\end{equation*}
from which one deduces the following estimate for
$\dot{\lambda}_{\gamma}$:
\begin{equation} \label{eq:4.19}
\mid \dot{\lambda}_{\gamma}(t, r) \mid \leq C(1 + r_{0} +
t)U_{\gamma}^{5}(t)
\end{equation}
Inserting (\ref{eq:4.16}), (\ref{eq:4.18}) and (\ref{eq:4.19}) in
the characteristic system, one has:
\begin{align*}
\mid \dot{v} \mid &\leq C(1 + r_{0} + t)U_{\gamma}^{6}(t) + C(1 +
r_{0} + t)U_{\gamma}^{3}(t)(1 + U_{\gamma}^{2}(t))\\
& \qquad + C(1 + r_{0} + t)^{3}Q_{\gamma}^{3/2}(t)
U_{\gamma}^{6}(t)\sqrt{1 + U_{\gamma}^{2}(t)}\\
& \qquad + C(1 + r_{0} + t) U_{\gamma}^{5}(t)\sqrt{1 +
U_{\gamma}^{2}(t)}\\
& \qquad + C(1 + r_{0} + t)^{3}Q_{\gamma}^{2}(t) U_{\gamma}^{6}(t)
\sqrt{1 + U_{\gamma}^{2}(t)}\\
& \qquad + C(1 + r_{0} + t)Q_{\gamma}^{1/2}(t) U_{\gamma}^{3}(t).
\end{align*}
Since
\begin{align*}
Q_{\gamma}^{1/2}(t) &\leq \sqrt{1 + Q_{\gamma}(t)} \leq 1 +
Q_{\gamma}(t),\\
Q_{\gamma}^{1/2}(t) &\leq (1 + Q_{\gamma}(t))^{3} \leq (1 +
Q_{\gamma}(t) + U_{\gamma}(t))^{3},
\end{align*}
one deduces from the above that
\begin{equation*}
\mid \dot{v} \mid \leq C(1 + r_{0} + t)^{3}(1 + Q_{\gamma}(t) +
U_{\gamma}(t))^{11}.
\end{equation*}
So, the integration of the above inequality on $[0, t]$ gives:
\begin{equation} \label{eq:4.20}
U_{\gamma}(t) \leq U_{0} + C\int_{0}^{t}(1 + r_{0} + s)^{3}(1 +
Q_{\gamma}(s) + U_{\gamma}(s))^{11}ds,
\end{equation}
where
\begin{equation*}
U_{0} := \sup \{ \mid v \mid \, | \, (\tilde{x}, v) \in \supp
\overset{\circ}{f} \}.
\end{equation*}
We recall that in the above, the constant $C$ depends on $q$ and
$\parallel \overset{\circ}{f} \parallel_{L^{\infty}}$ and does not
depend neither on $t$ and nor on $\gamma$. (\ref{eq:4.20}) shows
that we need an estimate for $Q_{\gamma}(t)$. We will proceed
exactly like we did in \cite{noundjeu2} when showing that the
sequence of iterates is bounded in the $L^{\infty}$-norm. So, we
differentiate $e^{-\lambda_{\gamma}}$ w.r.t $t$ and obtain:
\begin{equation*}
\left| \frac{\partial}{\partial t}e^{2\lambda_{\gamma}(t, r)}
\right| \leq 2\gamma Q_{\gamma}^{2}(t)\frac{\dot{m}_{\gamma}(t, r
)}{r}.
\end{equation*}
We see from the above inequality that an estimate for
$\frac{\dot{m}_{\gamma}}{r}$ is needed. Using the Gauss theorem
after insertion of the Vlasov equation, we deduce:
\begin{equation} \label{eq:4.21}
\begin{aligned}
\frac{\dot{m}_{\gamma}}{r} &= -4 \pi r e^{\mu_{\gamma} -
\lambda_{\gamma}}k_{\gamma} + \frac{1}{r}\int_{\mid y \mid \leq
r}\int_{\mathbb{R}^{3}}\frac{y . v}{\mid y \mid}(\mu_{\gamma}' -
\lambda_{\gamma}')e^{\mu_{\gamma} - \lambda_{\gamma}}f_{\gamma}
dvdy\\
&\quad - \frac{1}{r}\int_{\mid y \mid \leq
r}\int_{\mathbb{R}^{3}}\left( \dot{\lambda}_{\gamma}\sqrt{1 +
\gamma v^{2}} + \dot{\lambda}_{\gamma}\left( \frac{y . v}{\mid y
\mid} \right)^{2}\frac{1}{\sqrt{1 + \gamma v^{2}}} \right)
f_{\gamma} dvdy\\
&\quad - \frac{2}{r}\int_{\mid y \mid \leq r}\int_{\mathbb{R}^{3}}
e^{\mu_{\gamma} - \lambda_{\gamma}}\mu_{\gamma}'\frac{y . v}{\mid
y \mid}f_{\gamma}dvdy\\
&\qquad - \frac{q}{r}\int_{\mid y \mid \leq r}
\int_{\mathbb{R}^{3}}e^{\mu_{\gamma} + \lambda_{\gamma}}e_{\gamma}
\frac{y . v}{\sqrt{1 + \gamma v^{2}}}f_{\gamma}dvdy\\
&\qquad + \frac{2\pi}{r}\int_{0}^{r}s^{2}\frac{\partial}{\partial
t}(e^{2\lambda_{\gamma}}e_{\gamma}^{2})ds.
\end{aligned}
\end{equation}
Another part of the Maxwell equations is given by:
\begin{equation*}
\frac{\partial}{\partial t}(e^{\lambda_{\gamma}}e_{\gamma}) = -
\frac{q}{r}\gamma^{3/2}e^{\mu_{\gamma}}N_{\gamma},
\end{equation*}
where
\begin{equation*}
N_{\gamma}(t, r) = N_{\gamma}(t, \tilde{x}) :=
\int_{\mathbb{R}^{3}}\frac{\tilde{x} . v}{\sqrt{1 + \gamma
v^{2}}}f_{\gamma}(t, \tilde{x}, v)dv.
\end{equation*}
So, the last term in the right hand side of (\ref{eq:4.21})
yields:
\begin{equation*}
\frac{2\pi}{r}\int_{0}^{r}s^{2}\frac{\partial}{\partial
t}(e^{2\lambda_{\gamma}}e_{\gamma}^{2})ds = - 4\pi
\gamma^{3/2}\frac{q}{r}\int_{0}^{r}se^{\lambda_{\gamma} +
\mu_{\gamma}}e_{\gamma}N_{\gamma}ds,
\end{equation*}
and since
\begin{equation*}
\mid N_{\gamma}(t, r) \mid \leq
\begin{cases}
C(r_{0}/\sqrt{\gamma} + t/\gamma)U_{\gamma}^{3}(t)\\
C(r_{0} + t/\sqrt{\gamma})U_{\gamma}^{4}(t),
\end{cases}
\end{equation*}
we deduce the following estimate:
\begin{equation*}
\left| \frac{2\pi}{r}\int_{0}^{r}s^{2}\frac{\partial}{\partial
t}(e^{2\lambda_{\gamma}}e_{\gamma}^{2})ds \right| \leq C(1 + r_{0}
+ t)^{3}Q_{\gamma}^{1/2}(t)U_{\gamma}^{6}(t). \tag{4.21a}
\end{equation*}
Next, we give an estimate for the first term in the right hand
side of (\ref{eq:4.21}), after distinguishing the cases $r \leq
r_{0} + t/\sqrt{\gamma}$ and $r > r_{0} + t/\sqrt{\gamma}$:
\begin{equation*}
\mid -4\pi re^{\mu_{\gamma} - \lambda_{\gamma}}k_{\gamma} \mid
\leq C(1 + r_{0} + t)U_{\gamma}^{3}(t). \tag{4.21b}
\end{equation*}
Next, from (\ref{eq:4.1}), we deduce the following
\begin{equation*}
e^{\mu_{\gamma} - \lambda_{\gamma}}\lambda_{\gamma}' =
e^{\mu_{\gamma} + \lambda_{\gamma}}\left( -\gamma
\frac{m_{\gamma}}{r^{2}} + 4\pi \gamma r\rho_{\gamma} \right)
\end{equation*}
and then
\begin{equation*}
e^{\mu_{\gamma} - \lambda_{\gamma}}(\mu_{\gamma}' -
\lambda_{\gamma}') = e^{\mu_{\gamma} + \lambda_{\gamma}}\left(
2\gamma \frac{m_{\gamma}}{r^{2}} + 4\pi \gamma^{2} rp_{\gamma} -
4\pi \gamma r \rho_{\gamma} \right),
\end{equation*}
and using the above , one has:
\begin{align*}
\mid e^{\mu_{\gamma} - \lambda_{\gamma}}(\mu_{\gamma}' -
\lambda_{\gamma}') \mid (t, r) &\leq C(1 + r_{0} +
t)U_{\gamma}^{3}(t)\sqrt{1 + U_{\gamma}^{2}(t)}\\
&\qquad + C(1 + r_{0} + t)^{3} Q_{\gamma}^{3/2}(t)
U_{\gamma}^{6}(t)\\
&\qquad + C(1 + r_{0} + t)U_{\gamma}^{5}(t).
\end{align*}
So, with the help of this,and denoting by $A$ the second term in
the right hand side of (\ref{eq:4.21}), one deduces the following
estimate for $A$:
\begin{align*}
\mid A \mid &\leq C(1 + r_{0} + t)\left( r_{0} +
\frac{t}{\sqrt{\gamma}} \right)^{2}
U_{\gamma}^{11}(t)\\
&\qquad + C(1 + r_{0} + t)\left( r_{0} + \frac{t}{\sqrt{\gamma}}
\right)^{2}U_{\gamma}^{9}(t)\sqrt{1 + U_{\gamma}^{2}(t)}\\
&\qquad + C(1 + r_{0} + t)^{3} \left( r_{0} +
\frac{t}{\sqrt{\gamma}} \right)^{2}Q_{\gamma}^{3/2}
U_{\gamma}^{12}(t). \tag{4.21c}
\end{align*}
Using (\ref{eq:4.19}), we obtain the following estimate for the
third term, say $B$, in the right hand side of (\ref{eq:4.21}):
\begin{align*}
\mid B \mid &\leq C(1 + r_{0} + t)\left( r_{0} +
\frac{t}{\sqrt{\gamma}} \right)^{2}
U_{\gamma}^{8}(t)\sqrt{1 + U_{\gamma}^{2}(t)}\\
&\qquad + C(1 + r_{0} + t)\left( r_{0} + \frac{t}{\sqrt{\gamma}}
\right)^{2}U_{\gamma}^{10}(t). \tag{4.21d}
\end{align*}
Using (\ref{eq:4.18}), we deduce an estimate for the fourth term,
say $C$, in the right hand side of (\ref{eq:4.21}):
\begin{align*}
\mid C \mid &\leq C\gamma(1 + r_{0} + t)\left( r_{0} +
\frac{t}{\sqrt{\gamma}} \right)^{2}
U_{\gamma}^{7}(t)\sqrt{1 + U_{\gamma}^{2}(t)}\\
&\qquad + C\gamma(1 + r_{0} + t)^{3}\left( r_{0} +
\frac{t}{\sqrt{\gamma}} \right)^{2}Q_{\gamma}^{3/2}(t)
U_{\gamma}^{10}(t)\\
&\qquad + C\gamma(1 + r_{0} + t)\left( r_{0} +
\frac{t}{\sqrt{\gamma}} \right)^{2} U_{\gamma}^{9}(t)\\
&\qquad + C\gamma(1 + r_{0} + t)^{3}\left( r_{0} +
\frac{t}{\sqrt{\gamma}} \right)^{2}Q_{\gamma}^{2}(t)
U_{\gamma}^{10}(t). \tag{4.21e}
\end{align*}
Using (\ref{eq:4.15}), the fifth term, say $D$ in the right hand
side of (\ref{eq:4.21}) can be estimated by:
\begin{equation*}
\mid D \mid \leq C\gamma^{3/2}\left( r_{0} +
\frac{t}{\sqrt{\gamma}} \right)^{4}Q_{\gamma}^{1/2}(t)
U_{\gamma}^{7}(t). \tag{4.21f}
\end{equation*}
So, taking into account (4.21a), (4.21b), (4.21c), (4.21d),
(4.21e), (4.21f), one obtains the following estimate:
\begin{align*}
\left| \frac{\partial}{\partial t} e^{2\lambda_{\gamma}(t, r)}
\right| &\leq C(1 + r_{0} + t)^{3}Q_{\gamma}^{5/2}(t)
U_{\gamma}^{6}(t)\\
&\qquad + C(1 + r_{0} + t)Q_{\gamma}^{2}(t) U_{\gamma}^{3}(t)\\
&\qquad + C(1 + r_{0} + t)^{3}Q_{\gamma}^{2}(t)
U_{\gamma}^{9}(t)\sqrt{1 + U_{\gamma}^{2}(t)}\\
&\qquad + C(1 + r_{0} + t)^{5}Q_{\gamma}^{5/2}(t)
U_{\gamma}^{12}(t)\\
&\qquad + C(1 + r_{0} + t)^{3}Q_{\gamma}^{2}(t)
U_{\gamma}^{11}(t)\\
&\qquad C(1 + r_{0} + t)^{3}Q_{\gamma}^{2}(t) U_{\gamma}^{8}(t)
\sqrt{1 + U_{\gamma}^{2}(t)}\\
&\qquad + C(1 + r_{0} + t)^{3}Q_{\gamma}^{2}(t)
U_{\gamma}^{10}(t)\\
&\qquad + C(1 + r_{0} + t)^{3}Q_{\gamma}^{2}(t) U_{\gamma}^{7}(t)
\sqrt{1 + U_{\gamma}^{2}(t)}\\
&\qquad + C(1 + r_{0} + t)^{5}Q_{\gamma}^{5/2}(t)
U_{\gamma}^{10}(t)\\
&\qquad + C(1 + r_{0} + t)^{3}Q_{\gamma}^{2}(t)
U_{\gamma}^{9}(t)\\
&\qquad + C(1 + r_{0} + t)^{5}Q_{\gamma}^{4}(t)
U_{\gamma}^{10}(t)\\
&\qquad + C(1 + r_{0} + t)^{4}Q_{\gamma}^{5/2}(t)
U_{\gamma}^{7}(t),
\end{align*} thus,
\begin{equation*}
\left| \frac{\partial}{\partial t} e^{2\lambda_{\gamma}(t, r)}
\right| \leq C(1 + r_{0} + t)^{5}(1 + Q_{\gamma}(t) +
U_{\gamma}(t))^{17}.
\end{equation*}
So, the integration of the above inequality on $[0, t]$ yields:
\begin{equation*}
e^{2\lambda_{\gamma}(t, r)} \leq
e^{2\overset{\circ}{\lambda}_{\gamma}(r)} + C\int_{0}^{t}(1 +
r_{0} + s)^{5}(1 + Q_{\gamma}(s) + U_{\gamma}(s))^{17}ds
\end{equation*}
and bearing in mind that $(\overset{\circ}{f},
\overset{\circ}{\lambda}_{\gamma}, \overset{\circ}{e}_{\gamma})
\in D$, one has an estimate for $Q_{\gamma}$:
\begin{equation} \label{eq:4.22}
Q_{\gamma}(t) \leq C + C\int_{0}^{t}(1 + r_{0} + s)^{5}(1 +
Q_{\gamma}(s) + U_{\gamma}(s))^{17}ds,
\end{equation}
where $C$ is a constant which depends on $\overset{\circ}{f}$, $q$
and $\Lambda$, and not on $\gamma$ and $t$. Next, adding
(\ref{eq:4.20}) and (\ref{eq:4.22}), one has:
\begin{equation*}
Q_{\gamma}(t) + U_{\gamma}(t) \leq U_{0} + C + C\int_{0}^{t}(1 +
r_{0} + s)^{5}(1 + Q_{\gamma}(s) + U_{\gamma}(s))^{17}ds,
\end{equation*}
for $t \in [0, T_{\gamma}[$. Let $u :[0, T[ \rightarrow
\mathbb{R}_{+}$ be the maximal solution of the following integral
equation:
\begin{equation*}
u(t) = U_{0} + C + C\int_{0}^{t}(1 + r_{0} + s)^{5}(1 +
u(s))^{17}ds.
\end{equation*}
Then
\begin{equation*}
U_{\gamma}(t) \leq Q_{\gamma}(t) + U_{\gamma}(t) \leq u(t), \quad
t \in [0, T_{\gamma}[ \cap [0, T[
\end{equation*}
and with the continuation criterion proved in \cite{noundjeu2},
one concludes that $U_{\gamma}(t) \leq u(t)$ for $t \in [0, T[$
and this ends the proof of Theorem 4.1.

We now state and prove the essential result of this paper. This is
concerned with the convergence of a solution $(f_{\gamma},
\lambda_{\gamma}, \mu_{\gamma}, e_{\gamma})$ of $(EVM_{\gamma})$
to a solution $f$ for the (VP) system given by (\ref{eq:4.8}) and
(\ref{eq:4.9}), as $\gamma$ tends to $0$.
\begin{theorem} \label{T:4.2}
Let $0 < T \leq \infty$ be such that for every $\gamma \in [0, 1[$
and \\ $(\overset{\circ}{f}, \overset{\circ}{\lambda}_{\gamma},
\overset{\circ}{e}_{\gamma}) \in D$, the solution $(f_{\gamma},
\lambda_{\gamma}, \mu_{\gamma}, e_{\gamma})$ of $(EVM_{\gamma})$
exists on $[0, T[$ and
\begin{equation*}
f_{\gamma}(t, \tilde{x}, v) = 0, \, \mid v \mid > u(t), \,
\tilde{x} \in \mathbb{R}^{3}, \, t \in [0, T[,
\end{equation*}
where $u : [0, T[ \rightarrow \mathbb{R}_{+}$ is a continuous
function. Let $f \in C^{1}([0, + \infty[ \times \mathbb{R}^{6})$
be the solution of (VP) with $f(0) = \overset{\circ}{f}$. Then for
every $T' \in ]0, T[$ there exists a constant $C > 0$ such that
for any $\gamma \in ]0, 1]$ the following estimate holds:
\begin{equation} \label{eq:4.23}
\begin{aligned}
&\parallel f_{\gamma}(t) - f(t) \parallel_{L^{\infty}} + \parallel
\lambda_{\gamma}(t) \parallel_{L^{\infty}} + \parallel
\mu_{\gamma}(t) \parallel_{L^{\infty}} + \parallel
\dot{\lambda}_{\gamma}(t) \parallel_{L^{\infty}}\\
&\qquad + \parallel e_{\gamma}(t) \parallel_{L^{\infty}} + \left\|
\frac{1}{\gamma}\partial_{\tilde{x}} \mu_{\gamma}(t) - K(t) \right
\|_{L^{\infty}} \leq C\gamma,
\end{aligned}
\end{equation}
for every $t \in [0, T']$.
\end{theorem}
\textbf{Proof:} Take $0 < T' < T$. Let $C_{1}$ be the upper bound
of $u$ on $[0, T']$. Then one obtains, using the estimates above:
\begin{equation*}
U_{\gamma}(t), \, Q_{\gamma}(t), \, \frac{m_{\gamma}(t, r)}{r}, \,
\frac{m_{\gamma}(t, r)}{r^{2}} \leq C, \, t \in [0, T'], \, r > 0.
\end{equation*}
Using (\ref{eq:4.15}), we obtain
\begin{equation*}
\mid e_{\gamma}(t, r) \mid \leq C\gamma^{3/2}\left( r_{0} +
\frac{t}{\sqrt{\gamma}} \right) \leq C\gamma, \, r \geq 0, \, t
\in [0, T'], \, \gamma \in ]0, 1].
\end{equation*}
The remaining terms in (\ref{eq:4.23}) can be estimated exactly as
in the proof of Theorem 2 of \cite{rendall1}. But the sole change
is on estimate concerning $\frac{1}{\gamma}\partial_{\tilde{x}}
\mu_{\gamma}(t, \tilde{x}) - K(t, \tilde{x})$ that involves the
quantity $(m_{\gamma}/r^{2} - m/r^{2})(t, r)$. In our case, we
have after distinguishing the cases $r \leq r_{0} +
t/\sqrt{\gamma}$ and $r > r_{0} + t/\sqrt{\gamma}$:
\begin{equation} \label{eq:4.24}
\begin{aligned}
\left| \frac{m_{\gamma}}{r^{2}} - \frac{m}{r^{2}} \right|(t, r)
&\leq C\parallel f_{\gamma}(t) - f(t) \parallel_{L^{\infty}}\\
&\qquad + C\int_{0}^{r_{0} +
t/\sqrt{\gamma}}e^{2\lambda_{\gamma}(t, s)}e_{\gamma}^{2}(t, s)ds
\end{aligned}
\end{equation}
and using once again (\ref{eq:4.15}), the integral that appears in
(\ref{eq:4.24}) can be estimated as:
\begin{align*}
\int_{0}^{r_{0} + t/\sqrt{\gamma}}e^{2\lambda_{\gamma}(t,
s)}e_{\gamma}^{2}(t, s)ds &= q^{2}\gamma^{3}\int_{0}^{r_{0} +
t/\sqrt{\gamma}}\frac{1}{s^{4}}ds\left( \int_{0}^{s}\tau^{2}
e^{\lambda_{\gamma}(t, \tau)}M_{\gamma}(t, \tau)d\tau
\right)^{2}\\
&\qquad \leq C\gamma^{3}\int_{0}^{r_{0} + t/\sqrt{\gamma}}ds
\left( e^{\lambda_{\gamma}(t, \tau)}M_{\gamma}(t, \tau)d\tau
\right)^{2}\\
&\qquad \leq C\gamma^{3}Q_{\gamma}(t)U_{\gamma}^{6}(t)\left( r_{0}
+ \frac{t}{\sqrt{\gamma}} \right)^{3}\\
&\qquad \leq C\gamma.
\end{align*}
So, the proof of Theorem 4.2 is complete.

\end{document}